\documentclass[traditabstrac]{aa} 

\newcommand{\micron}{$\mu$m}

\usepackage{graphicx}
\usepackage{txfonts}
\usepackage{mathtools}  
\usepackage{colortbl}
\newcommand{\MJ}{$M_{\rm J}$~}

\usepackage{graphicx}
\usepackage{txfonts}
\usepackage{natbib}

\begin{document}

   \title{ALMA 870\,\micron \ continuum observations of HD 100546}
   \subtitle{Evidence of a giant planet on a wide orbit}
   \author{D. Fedele\inst{1,2},
            C. Toci\inst{3},
            L. Maud\inst{4},
            G. Lodato\inst{3}
          }

   \institute{INAF, Osservatorio Astrofisico di Arcetri, Largo Enrico Fermi 5, 50125, Firenze, Italy\\
              \email{davide.fedele@inaf.it}
         \and
         INAF, Osservatorio Astrofisico di Torino, Via Osservatorio 20, I-10025, Pino Torinese, Italy
         \and
         Dipartimento di Fisica, Universitá degli Studi di Milano, Via Celoria 16, Milano MI 20133, Italy
         \and  
         European Southern Observatory, Karl-Schwarzschild-Strasse 2, 85748, Garching, Germany
             }

   \date{...}

 
  \abstract
  {
 This paper reports on a new analysis of archival ALMA 870\,\micron ~dust continuum observations. Along with the previously observed bright inner ring ($r \sim 20-40\,$au), two addition substructures   are evident in the new continuum image: a wide dust gap, $r \sim 40-150\,$au, and a faint outer ring ranging from $r \sim 150\,$au to $r \sim 250\,$au and whose presence was formerly postulated in low-angular-resolution ALMA cycle 0 observations but never before observed. Notably, the dust emission of the outer ring is not homogeneous, and it shows two prominent azimuthal asymmetries that resemble an eccentric ring with eccentricity $e = 0.07 $.
  The characteristic double-ring dust structure of  HD 100546 is likely produced by the interaction of the disk with multiple giant protoplanets. This paper includes new smoothed-particle-hydrodynamic simulations with two giant protoplanets, one inside of the inner dust cavity and one in the dust gap. The simulations qualitatively reproduce the observations, and the final masses and orbital distances of the two planets in the simulations are 3.1\,\MJ at 15\,au and 8.5\MJ at 110\,au, respectively. The massive outer protoplanet substantially perturbs the disk surface density distribution and gas dynamics, producing multiple spiral arms both inward and outward of its orbit. This can explain the observed perturbed gas dynamics inward of 100\,au as revealed by ALMA observations of CO. 
  Finally, the reduced dust surface density in the $\sim 40-150\,$au dust gap can nicely clarify the origin of the previously detected H$_2$O gas and ice emission.  }
   \titlerunning{ALMA 870\,\micron \ continuum observations of  HD 100546}

   \maketitle
%

\section{Introduction}
Planet formation takes place in the interior of protoplanetary disks orbiting newly formed stars during the early stages of stellar evolution. Since its early observational campaigns, the Atacama Large Millimeter/submillimeter Array (ALMA) has revealed the internal structure of disks with unprecedented detail and sensitivity. In particular, the early ALMA observations brought to light the existence of disk substructures whose dust is largely confined in concentric rings separated by dust gaps \citep[e.g.,][]{Andrews16,Nomura16,Long18}. The origin of such gaps and rings are debated, but in some cases the large size and the pronounced depth of the dust gap can only be explained with the presence of giant protoplanets \citep[e.g.,][]{Isella16,Fedele17}. 
The unambiguous proof of the presence of planets and sub-stellar companions inside disks comes from direct imaging observations, as recently detected by \citet{Keppler18} and \citet{Ubeira20}.

The first directly imaged giant protoplanet candidate inside a disk was HD 100546 b, discovered by \citet[][later confirmed \citealt{Quanz15} and \citealt{Currie15}]{Quanz13}, who detected a bright point-like source at 3\,\micron \ using NaCo at the Very Large Telescope (European Southern Observatory, Chile) at nearly 70\,au from the central star. The authors estimated a mass of $\sim$ 15-20\,M$_{\rm Jup}$. Such a massive object should substantially perturb the dynamics and surface density of gas and dust, and these perturbations should be easily detectable by ALMA.

\smallskip

This paper presents a new analysis of archival ALMA observations of the 870\,\micron \ dust continuum emission of the HD 100546 disk, and it is structured as follows: The main stellar and disk properties are presented in Sect. \ref{sec:target};  Sect.~\ref{sec:observations} provides a description of the observations and data reduction method; the results and the comparison to near-infrared observations are described in Sect.~\ref{sec:results}; hydrodynamics simulations are presented in Sect.~\ref{sec:hydro}; and a discussion and conclusions are reported in Sects.~\ref{sec:discussion} and \ref{sec:conclusions}, respectively.

\begin{table*}
\caption{Observation details listing the EBs targeting HD 100546, antenna number, baseline ranges, and source information.}\label{tab:obs}      
\centering   
\footnotesize{
\begin{tabular}{l l l l l l l l l }      
\hline\hline       
Set & Project ID & EB & \#Ants  & Baseline  & Time on & Bandpass & Flux & Phase  \\ 
   &  &  &   & Range & source   & Cal. & Cal. & Cal.  \\ 
   &  &  &   & (m) & (min)  &  & &   \\ 
\hline                    
   A &  2015.1.00806 &    uid://A002/Xad5116/Xa70   &  36   &    17-10803   & 29.37 & J1427$-$4206 & J1107$-$4449 & J1147$-$6753 \\    
    &                &    uid://A002/Xad0ecf/X1ee8   &  44    &   17-14321  & 18.47 & J1427$-$4206 & J1107$-$4449 & J1147$-$6753  \\    
   B &  2016.1.00497 &    uid://A002/Xb9cc97/X8332   &  43     &   18-1124         & 35.63     & J1427-4206             & J1107$-$4449           & J1136$-$6827       \\    
    &                &    uid://A002/Xb9cc97/X86a1   &  43     &   18-1124         & 35.70    &  J1136-6827            & uses X8332              & J1136$-$6827        \\    
    &                &    uid://A002/Xc0015/Xe5c   &  45     &   15-1124         & 35.55    & J1427-4206          &   J1107$-$4449           & J1145$-$6954         \\    
    &                &    uid://A002/Xc0015/X16e3   &  43     &  15-1124         & 35.50     & J1427-4206             & J1107$-$4449           & J1145$-$6954 \\ 
 \hline                  
\end{tabular}
}
\end{table*}

\begin{figure*}
\centering
\includegraphics[width=8cm]{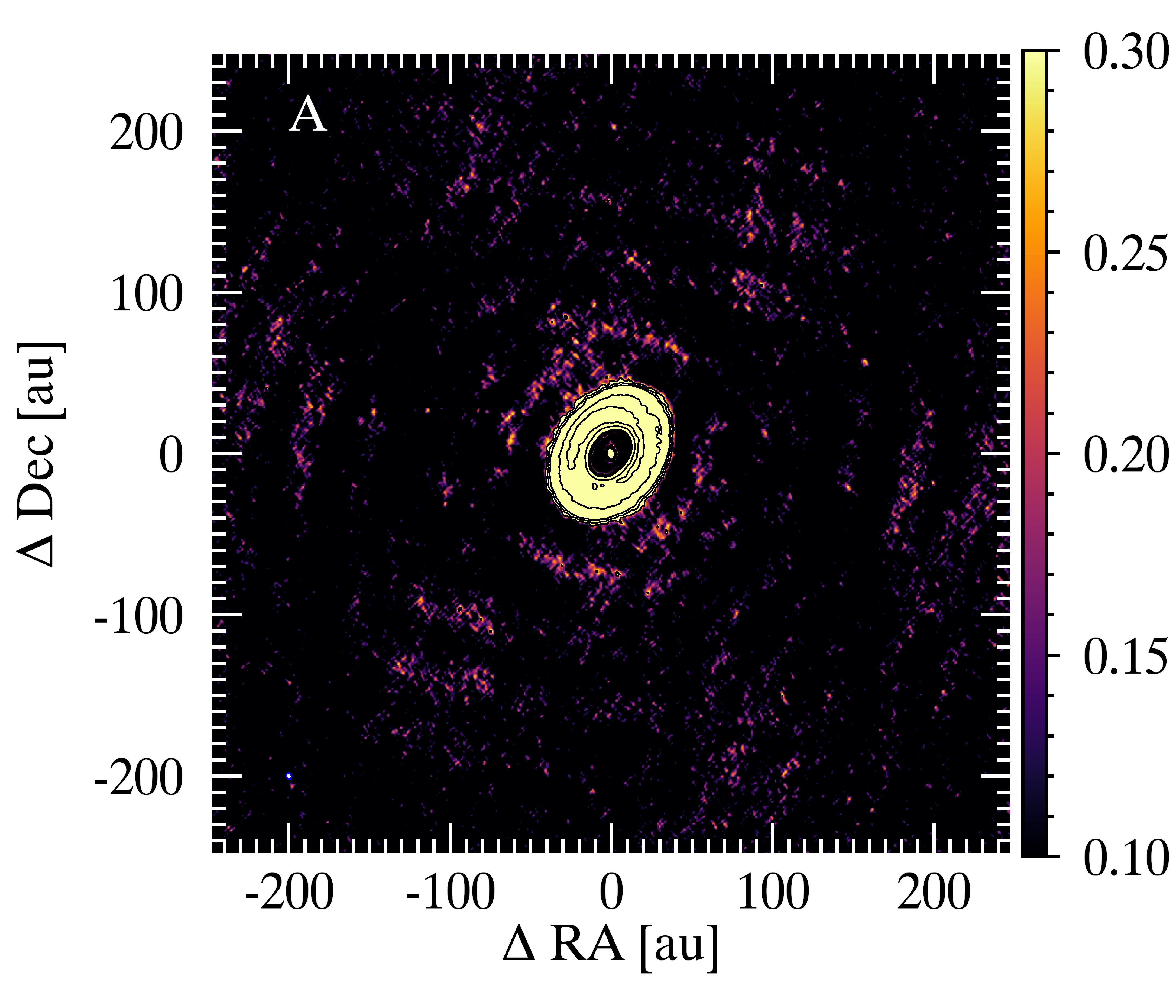}
\includegraphics[width=8cm]{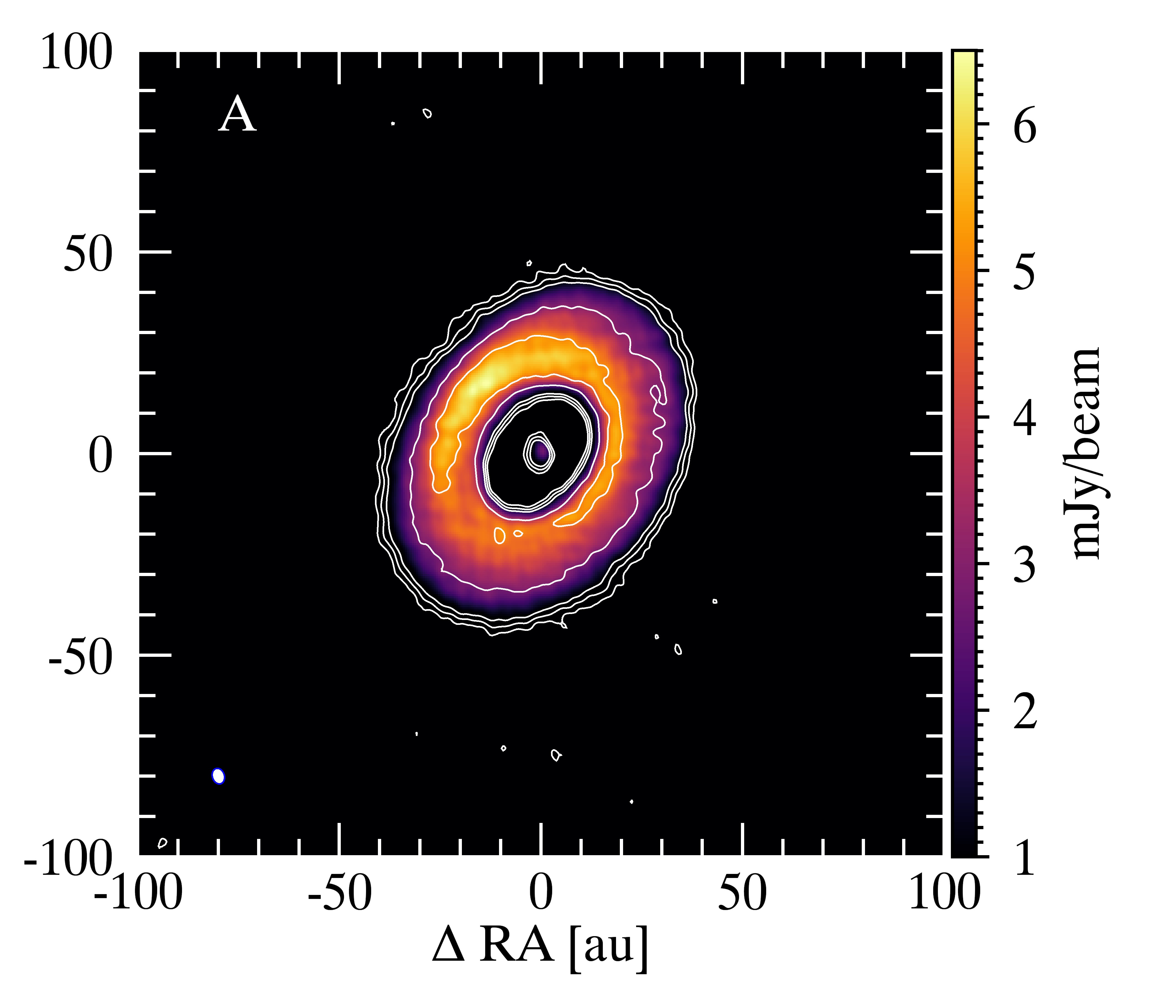}
\includegraphics[width=8cm]{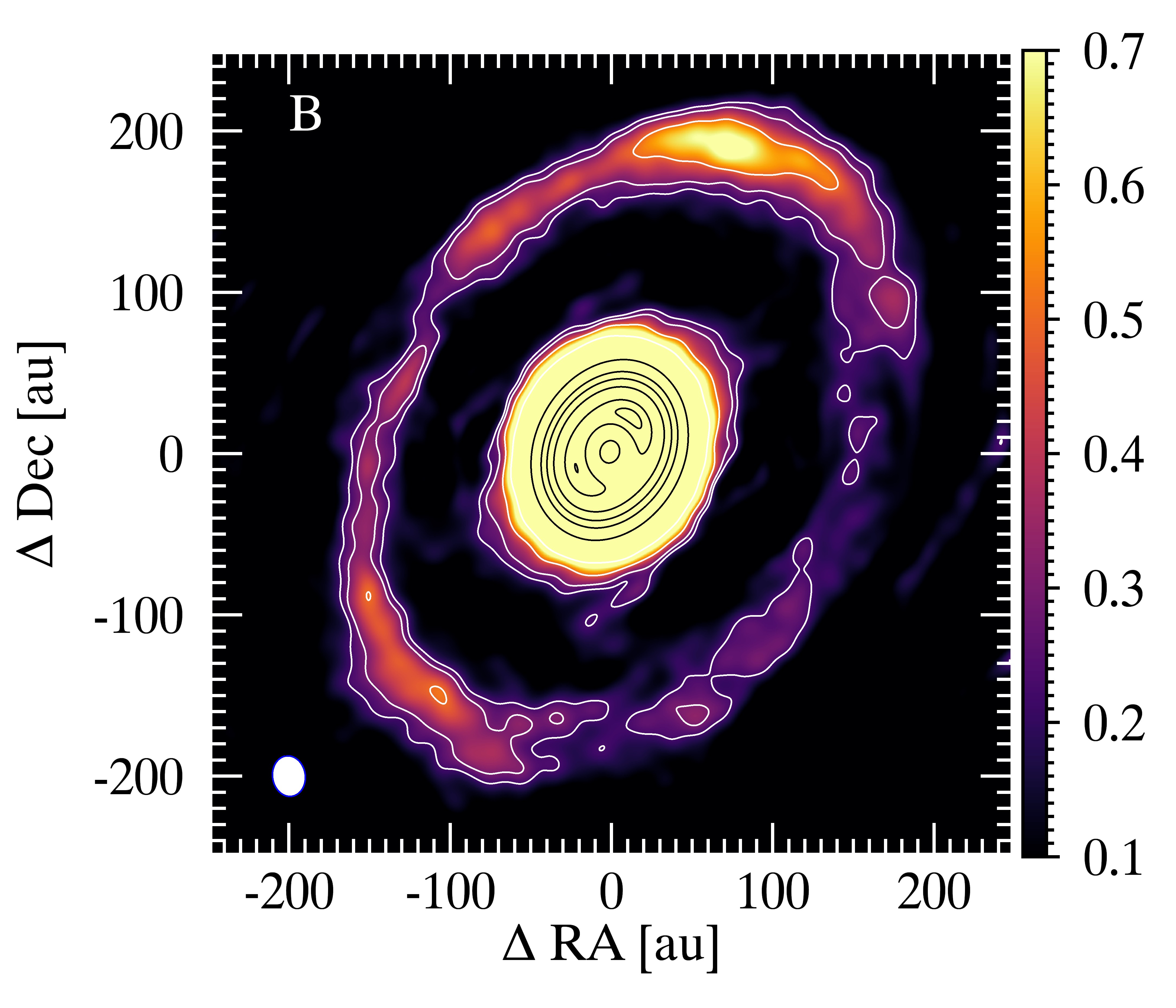}
\includegraphics[width=8cm]{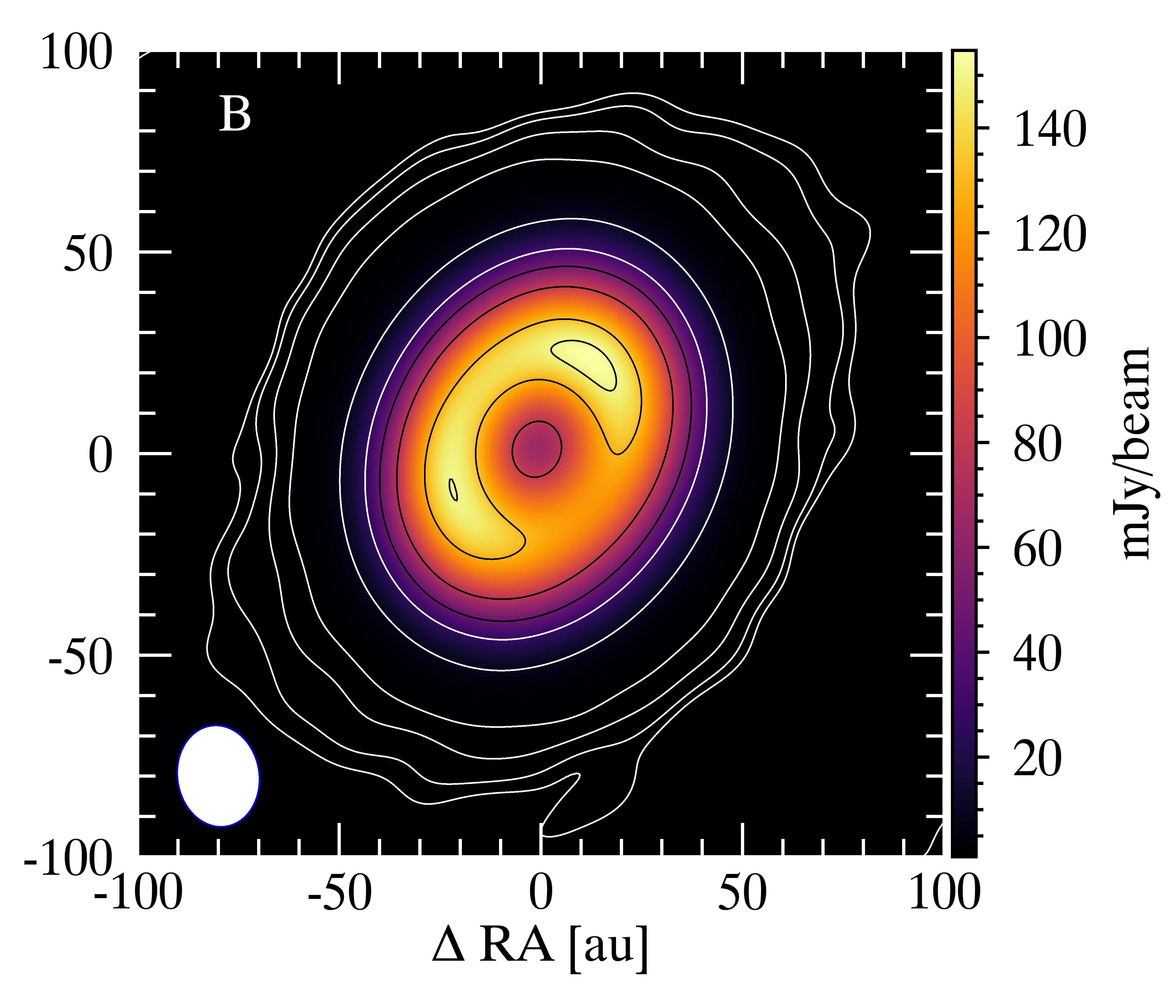}
\includegraphics[width=8cm]{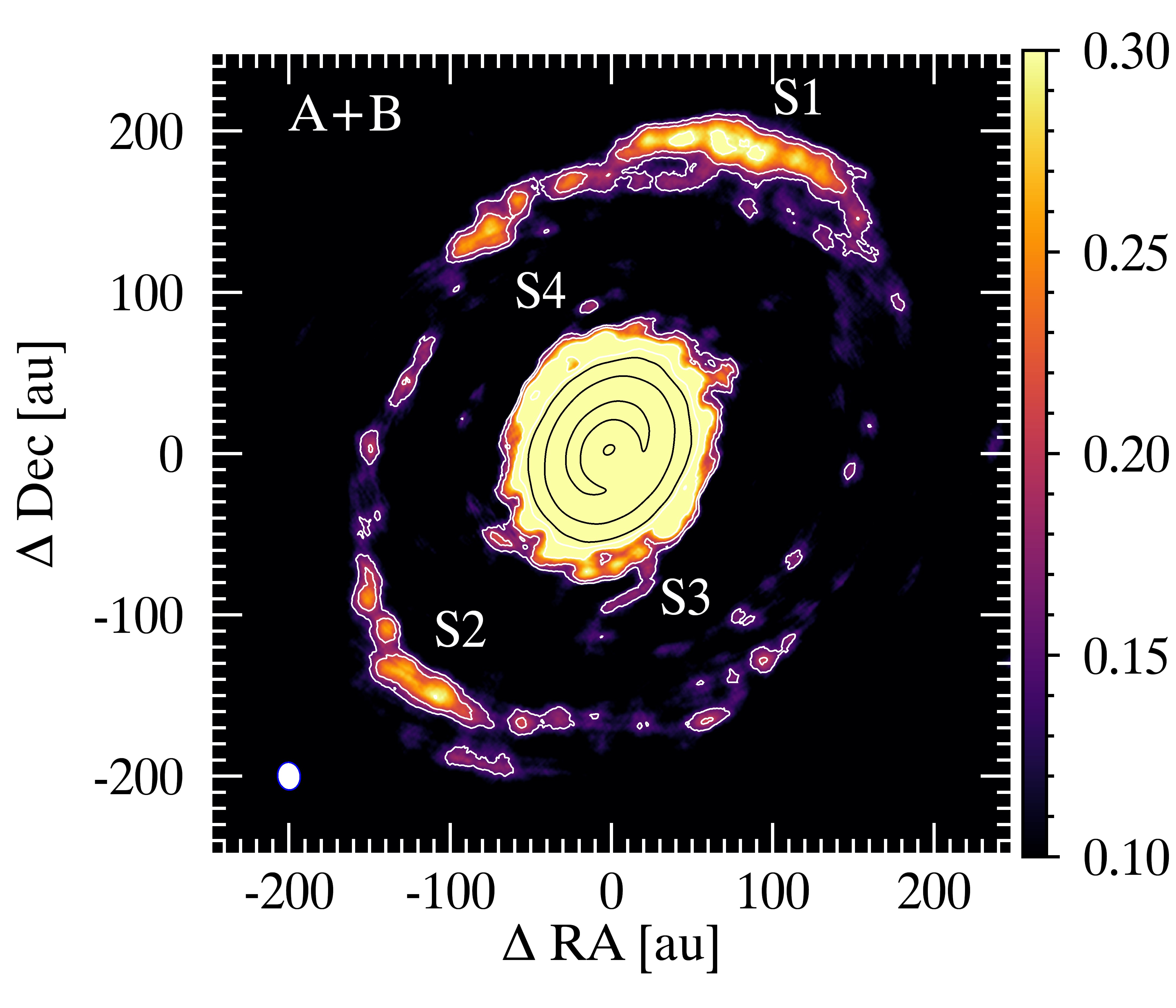}
\includegraphics[width=8cm]{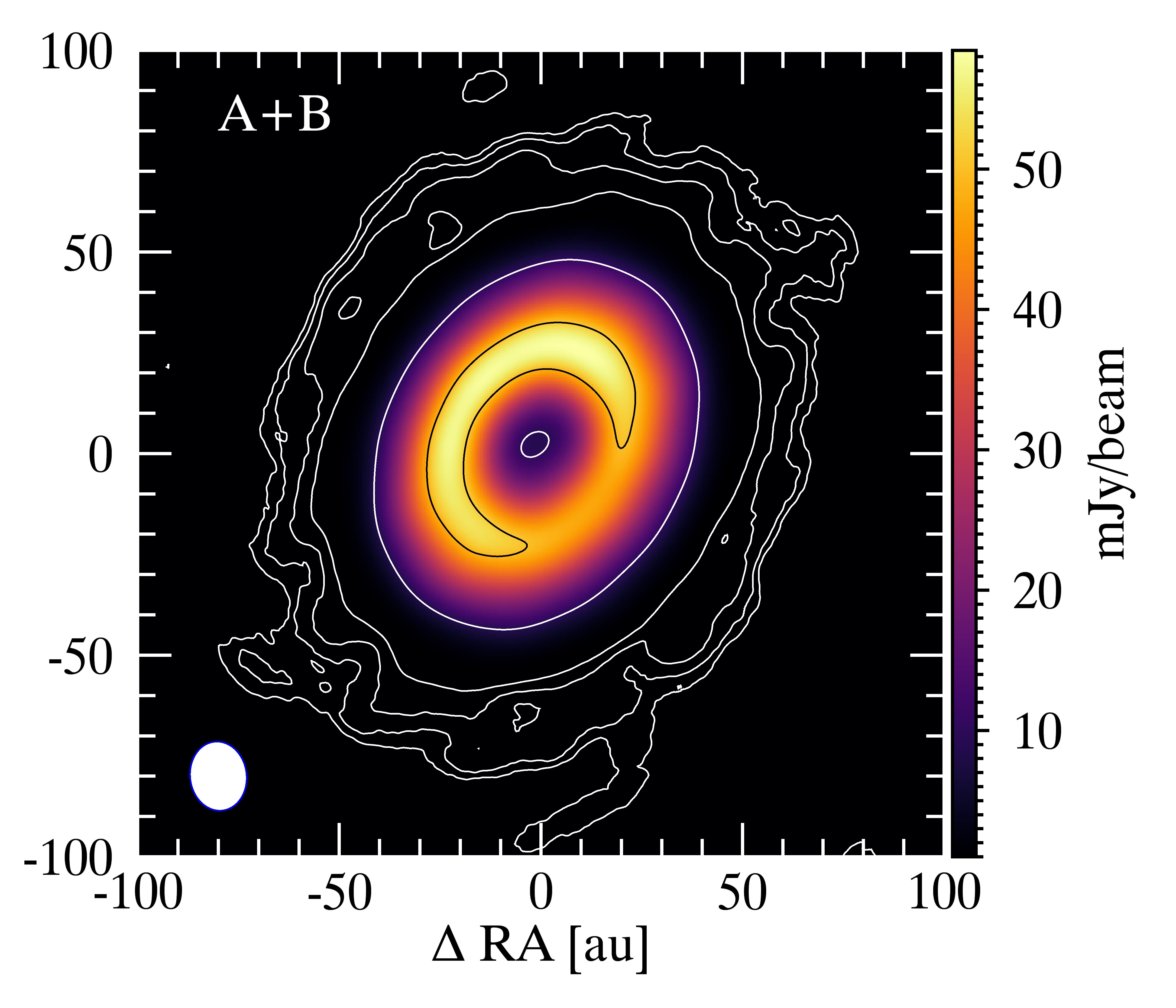}
\caption{ALMA band 7 continuum observations of HD 100546 from data set A ({\it top}), data set B ({\it middle}), and from the concatenated set ({\it bottom}).
The right panels show a zoom into the inner 100\,au.
The first overlaid contour in the bottom panels corresponds to the 4.5\,$\sigma$ level ($ 0.15$ mJy/beam). }
\label{fig:alma}
\end{figure*}

\section{The target}\label{sec:target}
HD 100546 is a 2.4 solar mass isolated pre-main-sequence star of spectral type B9 \citep[e.g.,][]{vdAncker98,Vioque18} located at a distance 
$d = 110.0\,pc \pm 0.6$ \citep{Brown18b}. The age of the star is poorly constrained; age values between 5 Myr and 10\,Myr are reported in the literature
\citep[e.g.,][]{vdAncker98,Vioque18}. Optical and near-infrared observations (molecular ro-vibrational and scattered light emission) have revealed an 
inner cavity of $r_{\rm cav} \sim 11-13\,$au \citep[e.g.,][]{Grady05,Brittain09,vanderplas09,Avenhaus14,fedele15}. The inner cavity is filled with (atomic) gas, as revealed by the presence of [\ion{O}{I}] 630\,nm emission \citep{Acke06}, whose origin is likely the photodissociation of OH molecules
\citep{Acke06}. High contrast imaging observations show the presence of multiple spiral arms at different spatial scales: Large-scale 
($> 2''$) spiral arms have been detected with the Hubble Space Telescope \citep[HST;][]{Ardila07} and with the Near-Infrared Coronagraphic Imager (NICI) at the GEMINI observatory  \citep{Boccaletti13}. A complex system of spiral arms has also been detected at small scales ($< 1''$) with extreme adaptive optics instruments,  such as the Spectro-Polarimetric High-contrast Exoplanet REsearch (SPHERE) at the VLT and the Gemini Planet Imager \citep[e.g.,][]{Avenhaus14, Garufi16, Follette17, Sissa18}.

Early ALMA cycle 0 observations by \citet{Walsh14} suggested a double ring structure of the dust continuum emission. More recent high-angular-resolution ALMA observations by \citet[][band 7]{Pineda19} and \citet[][band 6]{Perez20} resolved the disk at millimeter wavelengths, revealing a compact ring between $r \sim 20-40$\,au and confirming the cavity previously detected at  optical-to-near-infrared wavelengths.

ALMA also revealed asymmetric emission of CO on both small \citep{Pineda14, Walsh16, Pineda19} and large \citep{Miley19} spatial scales. Inward of 100\,au, the gas dynamics are perturbed: \citet{Walsh16} interpreted this perturbation as being due to a radial flow or to a warp on the inner disk. 
At larger spatial scales, \citet{Miley19} detected asymmetric emission spatially coincident with the structures (spiral arms) identified with HST.

\section{Observations and data reduction}\label{sec:observations}The data presented here are based on archival ALMA band 7 continuum observations from two different projects
(ID: 2015.1.00806, PI: Pineda, and ID: 2016.1.00497, PI: Pohl). Detailed information is given in Table~\ref{tab:obs}. 
The two projects cover a different range of spatial scales. 

The high-resolution observations from project 2015.1.00806 in the ALMA archive consist of one fully reduced execution block (EB) that used 36 antennas with baselines ranging from 17 to 10803\,m. These observations were previously published in \citet{Pineda19}, along with an additional second EB. This second EB, which uses 44 antennas with baselines ranging from 17\,m to 14321\,m, was classed as ``semi-pass'' data due to not being a fully complete observation, achieving only 18 minutes of an expected 29 minutes of on-source time for HD 100546. The ALMA quality assessment report indicated that these data are otherwise good. A calibration script was generated for the second EB, and both data sets were recalibrated using \textsc{casa} version 5.4.0 \citep{Mcmullin07}. The EBs were combined for imaging and subsequent phase and amplitude self-calibration. Phase-only self-calibration was conducted in an iterative manner starting from solution intervals of the length of the target scan time ($\sim$54\,s) and decreasing down to 6\,s. Thereafter, one round of amplitude self-calibration was performed using the scan time solution interval.  

With the aim of increasing the sensitivity toward the large angular scales, four EBs from the most extended array configuration ($\sim$15-1124\,m) taken as part of the second project, 2016.1.00497, were recalibrated using the archive reduction scripts in \textsc{casa} version 5.4.0. 
Data from this project have not previously been published and were obtained in full polarization mode, although only the total intensity is used in this work. Self-calibration was performed in stages. Half of the EBs were taken in 2016 and the other half in 2017, and thus they required independent phase center alignment. This was achieved by adopting the high-resolution observations  (data set A) as a reference model to form (long-term) phase solutions, which only act to shift the phase center to that of the reference model. Thereafter, self-calibration was conducted on the four combined EBs of this project as described above, with the exception that the initial solution had a scan time solution interval of $\sim$300\,s. 

Finally, the self-calibrated concatenated visibilities were imaged with \textsc{casa.tclean}. The image parameters are listed in Table~\ref{tab:log}. Channels containing line emission were excluded from the continuum images. Figure~\ref{fig:alma} shows the final continuum images for the two data sets (A and B) imaged independently and with the data sets combined (A+B). All images used a Briggs weighting, and robust = 0.5. The synthesized beam of the merged data set (A+B) is $0\farcs156~\times~0\farcs127$ (17 $\times$ 14\,au) with a position angle of 
7$^{\circ}$, and the achieved r.m.s. is 0.033 mJy beam$^{-1}$.

\begin{figure*}
\centering
\includegraphics[width=9cm]{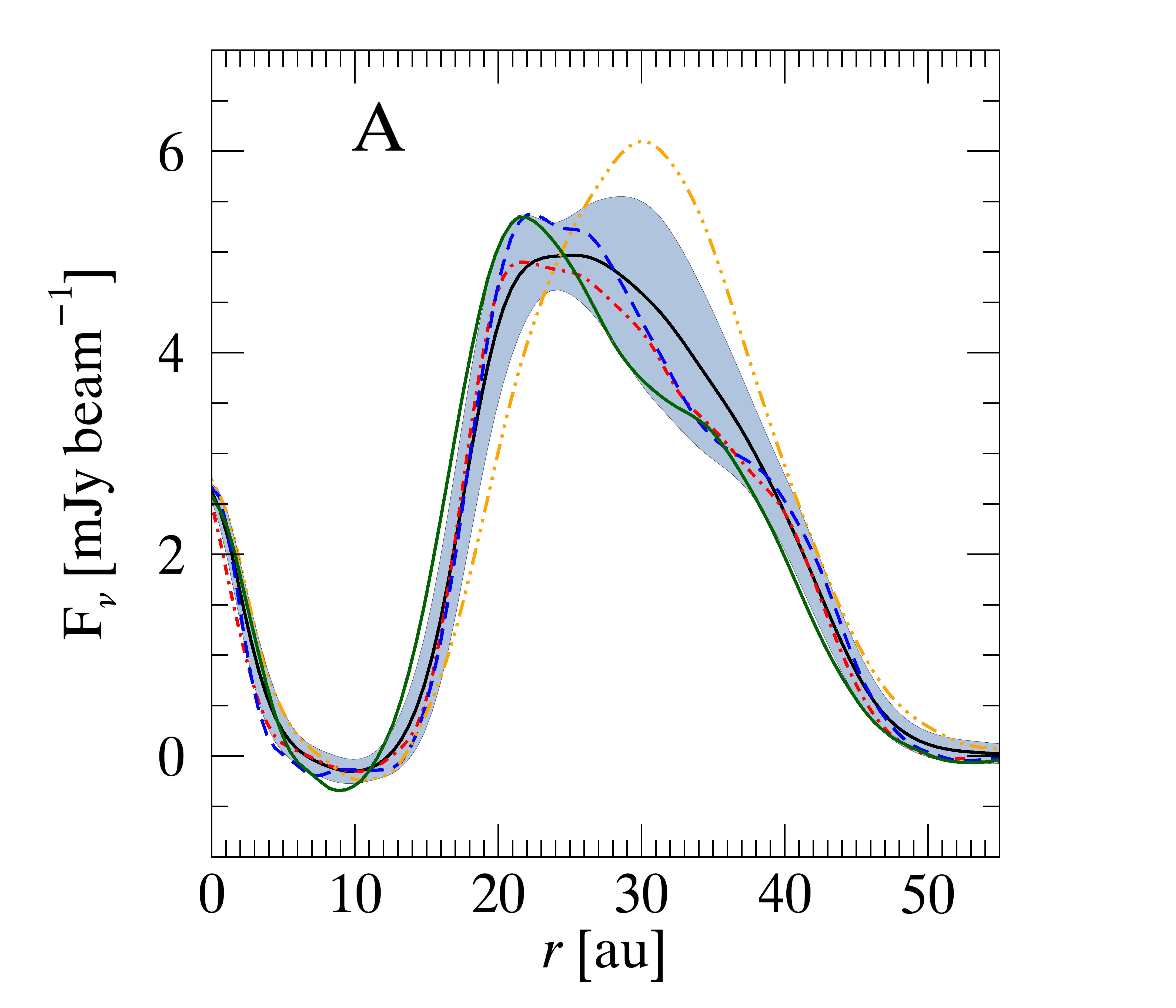}
\includegraphics[width=9cm]{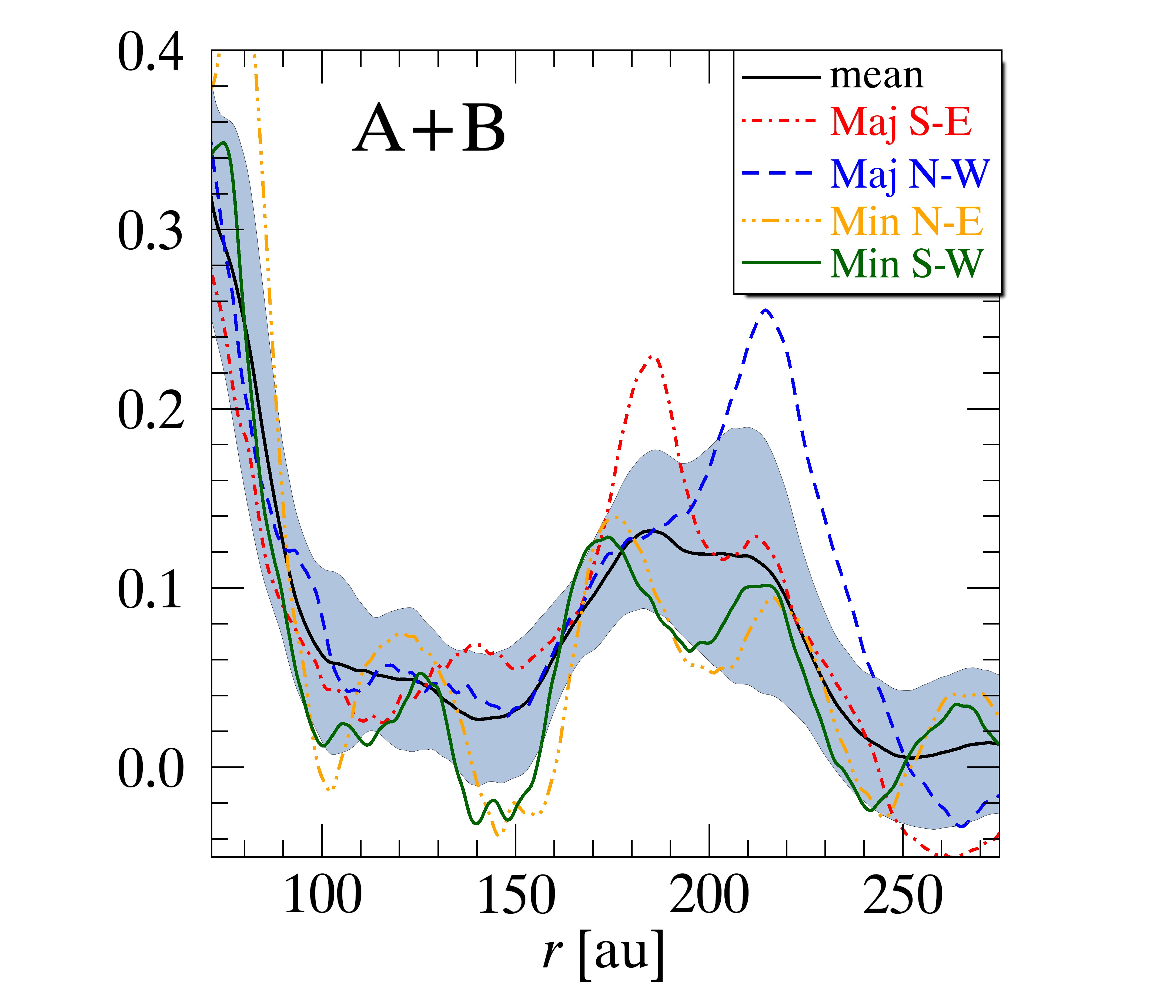}
\caption{Surface brightness profiles of the inner (left, from data set A) and outer (right, combined data sets A+B) disk of 
HD 100546. The black curve and the gray region represent the azimuthally averaged profile and standard deviation obtained after
de-projecting the images by 42$^{\circ}$. The profiles along the semimajor and semiminor axes are overplotted in color; they were obtained by averaging the radial profiles over a range of $\pm$ 10$^{\circ}$ around the corresponding axis.}
\label{fig:cut}
\end{figure*}

\section{Results}\label{sec:results}
\subsection{Image inspection}
The ALMA band 7 continuum emission of HD 100546 shows two prominent structures: an inner ring located between $r \sim 20 $ and $ 40\,$au and 
an outer ring-like structure between $r \sim 150$ and 250\,au. A gap between $r \sim 40$ and $ 150$\,au separates the two rings. 
The properties of the inner ring have already been presented in \citet{Pineda19} and \citet{Perez20}.
The merged data set allows us to detect and spatially resolve the outer ring that was postulated by \citet{Walsh14} based on 
low-resolution ($\sim 0\farcs95 \times 0\farcs42$) ALMA cycle 0 data. It should be noted that the outer ring is much fainter than the inner one, with 
a brightness ratio of $\sim 10^{-3}$ between the two. Notably, the outer ring is not homogeneous: As outlined in Fig.~\ref{fig:alma} 
(bottom left panel), the continuum emission is stronger in two arm-like structures in the northwestern (substructure ``S1'') and southeastern (``S2'') direction, respectively. 
In the southwestern direction, there is evidence of an additional arm (``S3'') inward of the 
main ring. Finally, a point-like source is detected at a 5\,$\sigma$ level at $r \sim 106\,$au and a position angle of $\theta \sim 10^{\circ}$ (``S4''). Notably, the position of S4 is right in the middle of the dust gap. The nature of S4 is unclear as its position is symmetrical to that of S3. 
From the inspection of the point spread function (PSF) we can rule out S3 being an artificial feature inherited from the PSF. Nevertheless, the presence of low level flux in its proximity may suggest that S4 is part of a spatially extended emission, such as a spiral arm. 

\subsection{Geometrical properties}
The geometrical properties of the disk were estimated by fitting  continuum emission with \textsc{galario} \citep{Tazzari18} and 
\textsc{emcee} \citep{Foreman13}. The underlying model of the brightness profile is the sum of two Gaussian rings. The inclination, 
position angle, and center of the two rings are free parameters. 
The \textsc{galario} fit results are reported in the appendix.
The best-fit model gives an inclination of 41.7$^{\circ}$ and a position angle of 146$^{\circ}$, in very good agreement with those estimated by \citet[][$i=42^{\circ} \pm 5^{\circ}$, PA =$145^{\circ} \pm 5^{\circ}$]{Ardila07} 
based on the HST observations. This implies that the inner ring and the outer disk (traced by the HST images) have the same inclination and position angle.

Figure~\ref{fig:cut} shows the brightness profiles after de-projecting for the disk inclination. The figure shows the azimuthally
averaged profile and the projection along the semi-axis (averaged over a range of $\pm$ 10$^{\circ}$). The inner disk (left panel, 
from data set A) is asymmetric, as previously noted by \citet{Pineda19} and \citet{Perez20}, with a sharp inner edge at around $r\sim 
20$\,au and a smoother outer edge extending from nearly $30$\,au to $50$\,au. Along the minor axis, the ring is brighter in the northeast direction, peaking at $r\sim 31\,$au. On the contrary, the dust emission along the other semi-axis peaks at $r \sim 20-25\,$au.
The outer disk (right panel, from data set A+B) shows multiple features: a narrow ring at $r \sim 80\,$au and 
an azimuthal asymmetry in the outer ring, with the emission peaking at different radii along the different projections. 
We note in particular that the emission peaks at $r \sim 180\,$au along the southeastern semimajor axis projection (146$^{\circ}$) 
and at $r \sim 220\,$au along the northwestern semimajor axis ( 326$^{\circ}$). These two peaks correspond to the substructures S1 and S2 
in Fig.~\ref{fig:alma}. 

\begin{table}
\caption{Image details.}             
\label{tab:log}      
\centering          
\begin{tabular}{l l l l}     
\hline\hline       
Set & Project ID & Beam (PA) & r.m.s. \\ 
       &         &             &  [mJy/beam] \\
\hline                    
   A &  2015.1.00806 & 0\farcs036 $\times$ 0\farcs027 (18) & 0.062 \\    
   B &  2016.1.00497 & 0\farcs229 $\times$ 0\farcs184 (8)   & 0.045 \\    
 A+B &                      & 0\farcs156 $\times$ 0\farcs127 (7)   & 0.033 \\    
\hline                  
\end{tabular}
\end{table}

\begin{figure*}
\centering
\includegraphics[width=9cm]{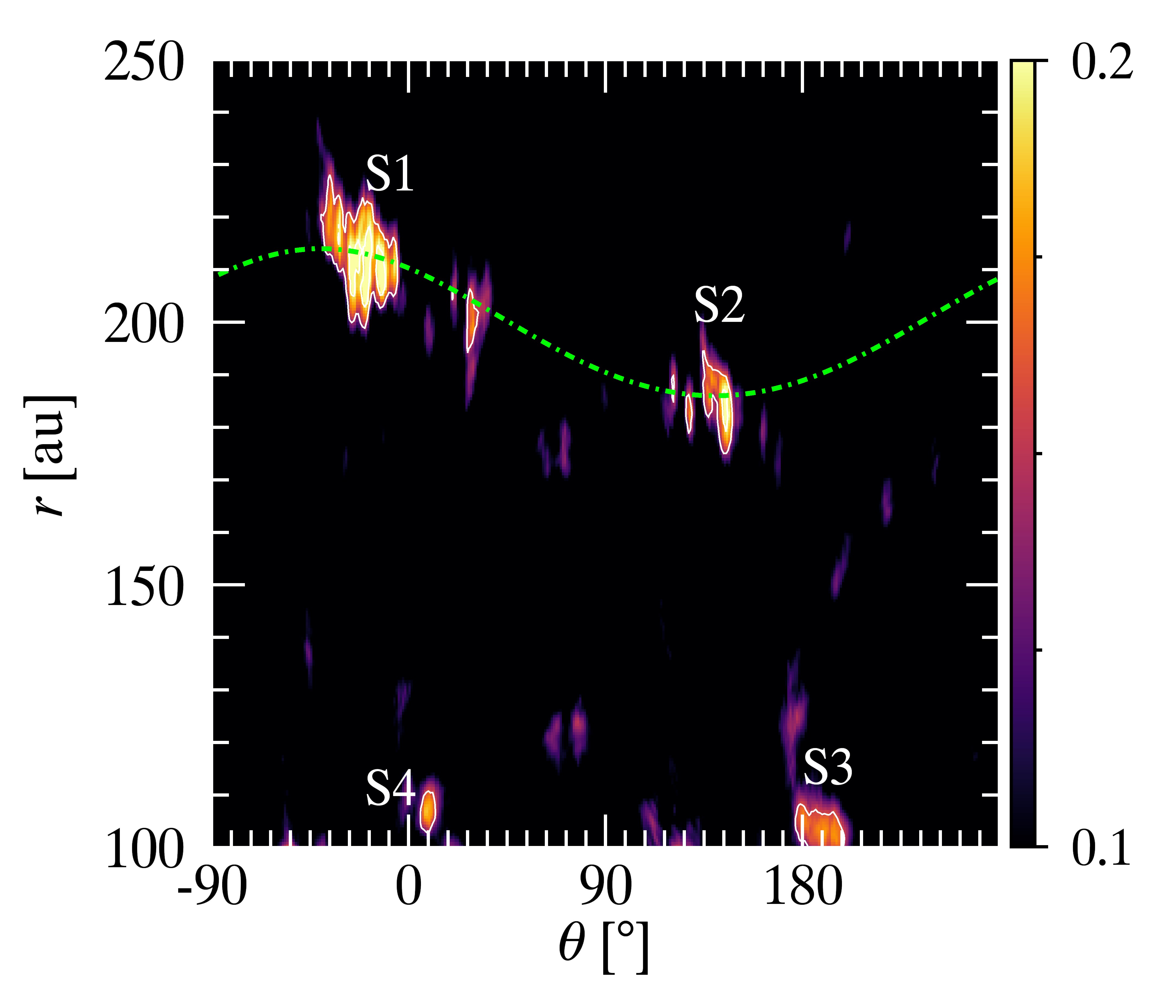}
\includegraphics[width=9cm]{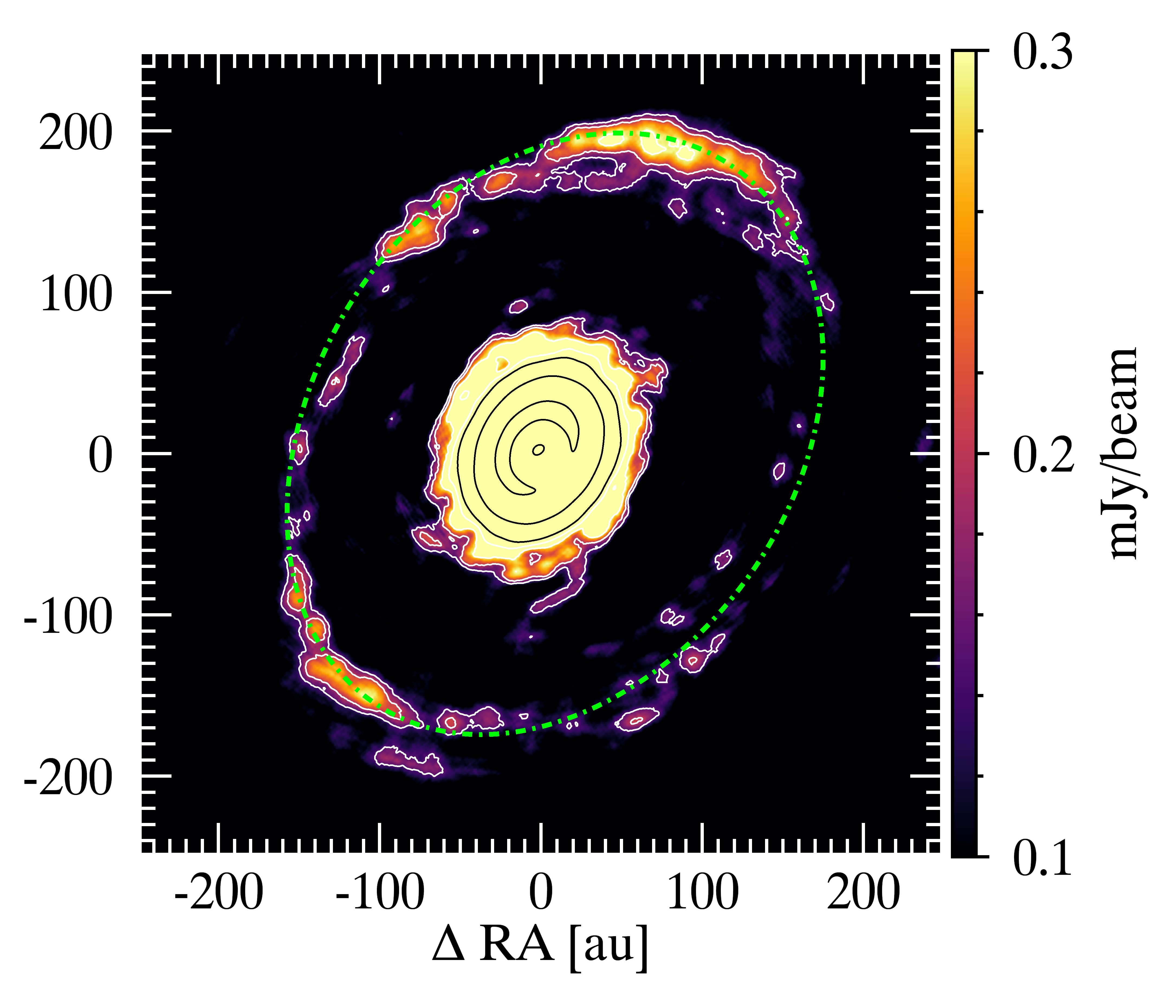}
\caption{Eccentricity of the outer ring. ({\it left}) Surface brightness of the outer disk region in polar coordinates after de-projection and subtraction of a Gaussian 
ring centered at $r=200\,$au. The dot-dashed green line is the best fit of the outer ring with an eccentricity of 0.07 (see Sect.~\ref{sec:results} 
for details). ({\it right}) Surface brightness with the eccentric ring overlaid.}
\label{fig:polar}
\end{figure*}
   
\subsection{Eccentric ring}
The azimuthal asymmetries outlined above point to an eccentric outer ring.
The left panel of Fig. \ref{fig:polar} shows the residual of the band 7 continuum image in polar coordinates after de-projection and subtraction of a Gaussian ring centered at $r=199.5\,$au with a width of 23.32\,au (see the \textsc{Galario} fit results in the appendix). Residuals are present, corresponding to S1 and S2. Fitting the position of these residuals with an eccentric ring gives an eccentricity $e = 0.07\pm0.01$. 
The right panel of Fig.~\ref{fig:polar} shows the best-fit eccentric ring overlaid on the surface brightness profile.

\subsection{Comparison to near-infrared observations}\label{sec:nir}
\begin{figure*}[!t]
   \centering
   \includegraphics[width=9cm]{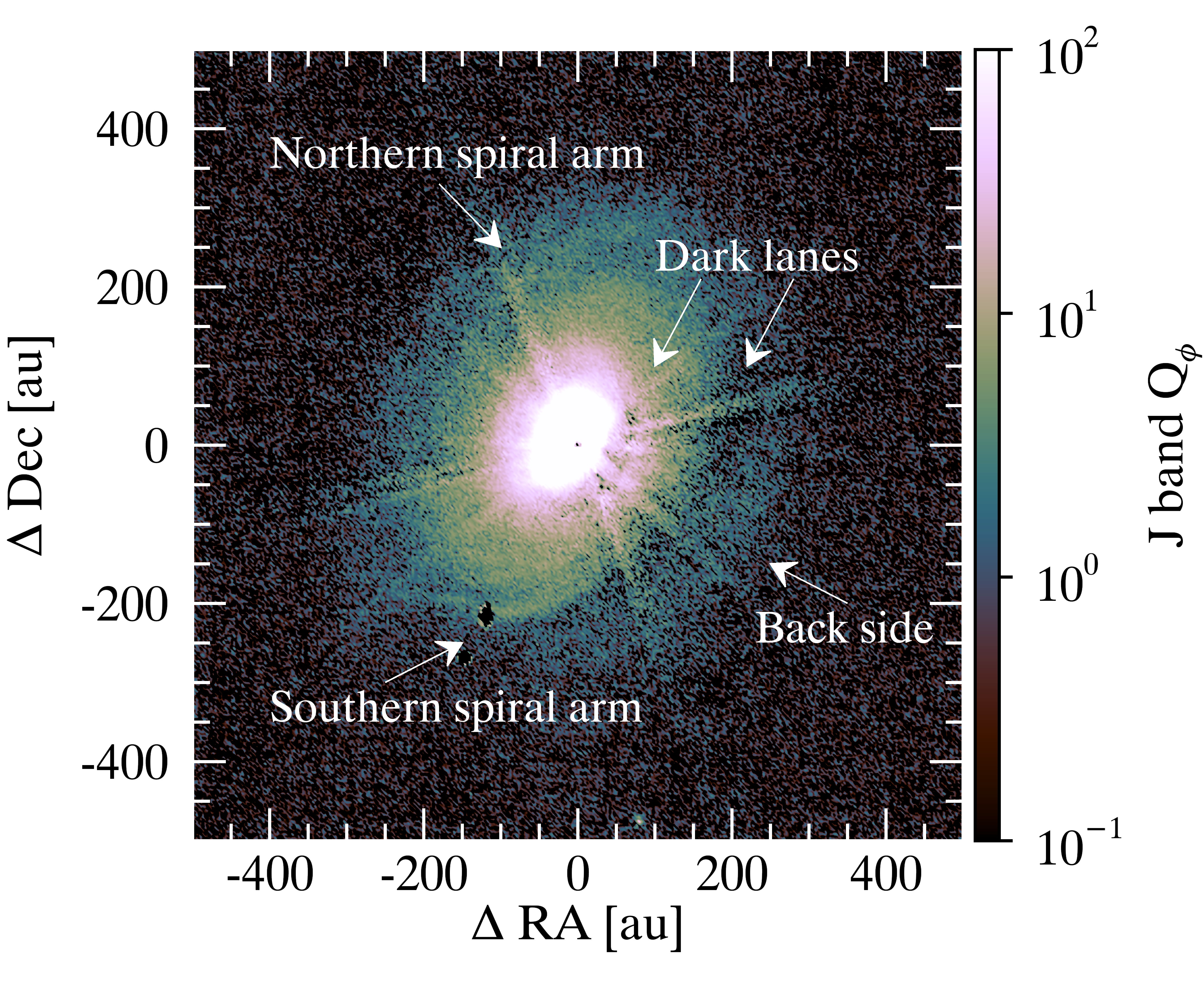}
   \includegraphics[width=9cm]{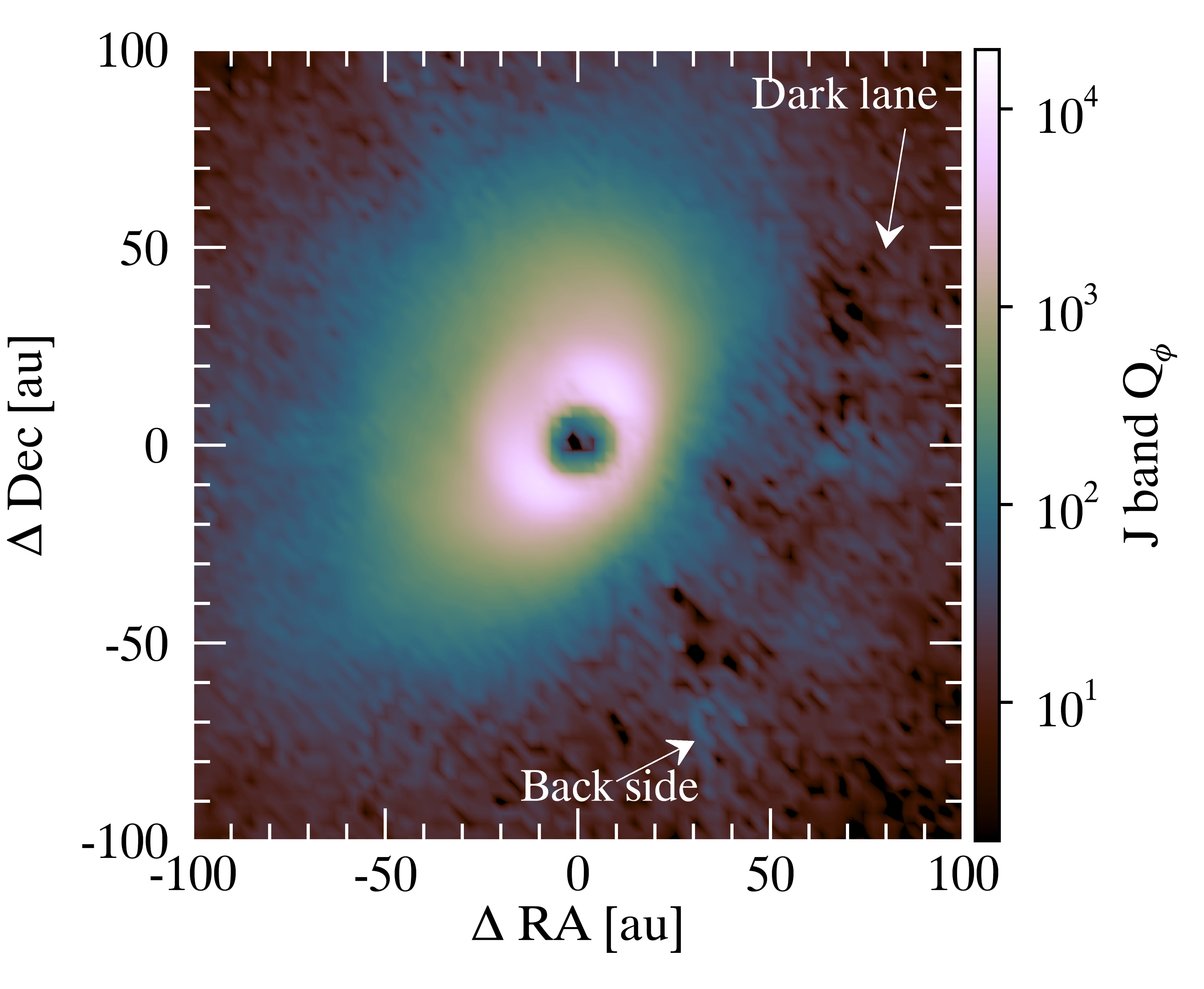}
   \includegraphics[width=9cm]{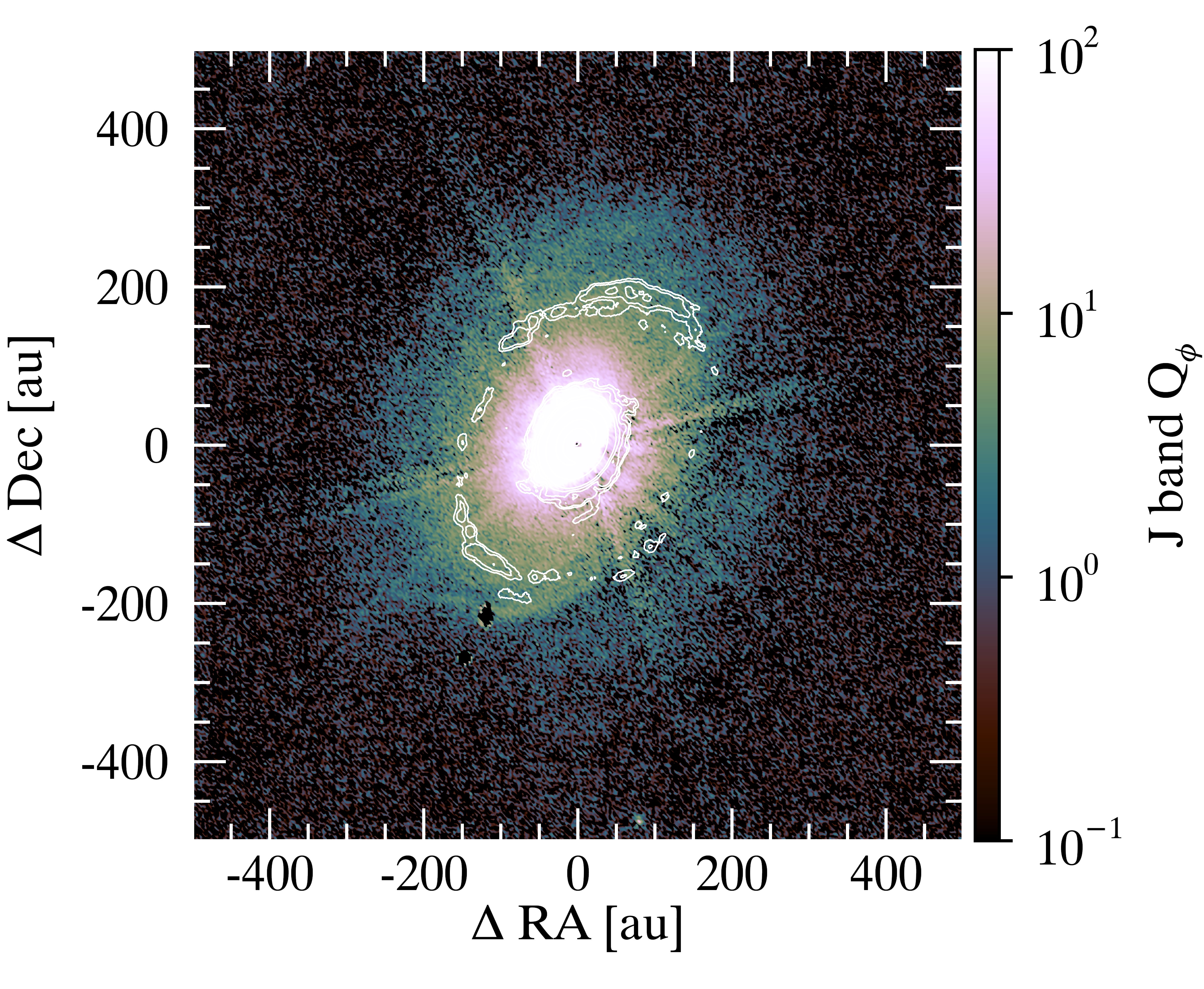}
   \includegraphics[width=9cm]{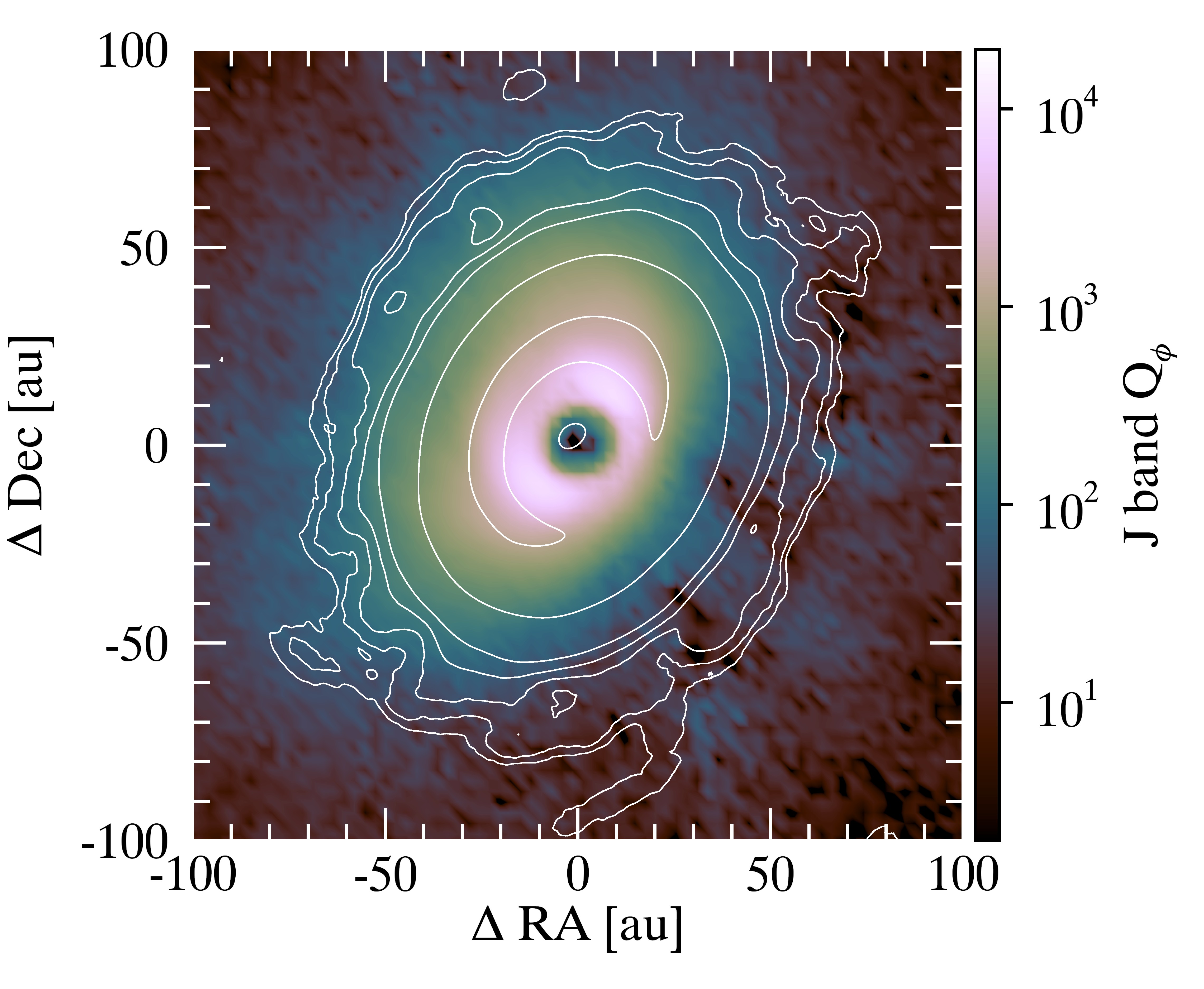}
    \caption{VLT/SPHERE IRDIS polarimetric observations of HD 100546 (from \citealt{Sissa18}) showing the large-scale (left) and small-scale (right) structures. The main substructures are overlaid in the top panels.  The ALMA 870\,\micron ~contours are overlaid in the bottom panels.}
         \label{fig:hst}
   \end{figure*}

Figure~\ref{fig:hst} shows the near-infrared polarimetric observation of HD 100546 taken with SPHERE/IRDIS (the InfraRed Dual-band Imager and Spectrograph) at the VLT (J band Q$_{\phi}$ from \citealt{Sissa18}). The main substructures are indicated: The two large-scale spiral arms, 
two dark lanes, and the disk back side are clearly visible \citep[e.g.,][]{Ardila07, Boccaletti13, Avenhaus14, Garufi16,Sissa18}. 
The ALMA 870\micron ~continuum emission is overlaid in the bottom panels of Fig.~\ref{fig:hst}: At near-infrared wavelengths, the disk appears radially more extended than at millimeter wavelengths. 
The inner dark lane is spatially coincident with the ALMA inner ring. This implies that the surface density of the micron-sized particles is partly reduced beyond the inner ring. 
The near-infrared emission extends inward down to $r \sim 11-13$\,au \citep[see also][]{Avenhaus14,Garufi16}, while the millimeter emission rises beyond 15\,au and the emission peak is at $r \sim 28\,$au \citep[as previously noted by][]{Pineda14}. We note that the ro-vibrational emission of CO \citep[e.g.,][]{Brittain09,vanderplas09,Hein14} and OH \citep[e.g.,][]{Brittain14, fedele15} also extends inward down to $r \sim 11-13$\,au.

\subsection{Planet-induced perturbations}
The differences in the near-infrared and millimeter surface brightness are likely the outcome of dynamical disk-planet interaction: The large dust grains traced by ALMA are settled to the disk midplane and trapped in two main ring-like structures that are produced by two embedded giant protoplanets; the small dust grains traced by the J band polarimetric observations are  dynamically coupled to the gas, which is more extended than the large dust grains both in the vertical and radial direction and diffuses inside the gap \citep[e.g.,][]{Zhu12}.
The sharp edge of the inner dust ring at 20\,au also implies the presence of a massive inner planet that is needed to stop the inward migration of millimeter dust, as already suggested by several authors (see Sect.~\ref{sec:target}). The presence of an outer massive planet is consistent with: 1) the outer dust gap, 2) the perturbed dynamics of the gas inward of 100\,au (\citealt{Pineda14} and \citealt{Walsh17a}), and 3) the multiple spiral patterns observed in the near-infrared. All of these findings are in agreement with theoretical calculations of disk-planet interactions \citep[e.g.,][]{GoldreichTremaine1979,OgilvieLubow2002,Crida07,Dipierro2017}.

\smallskip
 
Following \citet{Rosotti16} and \citet{Lodato2019}, in the case of a gap opened by a planet, the dust gap width ($\Delta_{\rm p}$) scales with the Hill radius of the planet as $\Delta_{\rm p}=\simeq 5.5 $R$_{\rm H}$, where R$_{\rm H}$ depends on the planet ($M_{\rm p}$) and stellar ($M_{\star}$) masses as $[M_{\rm p}/(3M_{\star})]^{1/3}$a$_{\rm p}$. The estimated planetary mass is $\sim$ 5.5 - 6\,M$_{\rm J}$ for a $\Delta_p =100$ au and an orbital radius of 100 and 110\,au, respectively.

\section{Comparison to hydrodynamical simulations}\label{sec:hydro}
To answer the question of whether a single massive planet at 100-110\,au is capable of producing the structures in the outer disk described above, we carried out a number of 3D smoothed-particle-hydrodynamic (SPH) simulations using the PHANTOM code \citep{Price18} and changing the initial conditions in terms of disk mass and size, surface density profile, and planet eccentricity. The following sections present the results of the simulation that qualitatively reproduces the observational findings best.

\subsection{Physical setup}
To simulate the dynamics and evolution of a disk composed of dust and gas, the multigrain one-fluid method \citep{Hutchison18} was adopted. This method is best suited for small values of the Stokes number ($St < 1$). Smoothed-particle-hydrodynamic particles represent both gas and dust and are evolved with an algorithm based on the terminal velocity approximation \citep{Laibe14}. Back-reaction from the dust onto the gas is naturally included in the code. The system was modeled with a central star and two giant planets embedded in the disk, which were allowed to migrate and accrete mass. The fluid self-gravity was neglected.

The disk was set up as in \citet{Toci20}: The central star was surrounded by a disk of $10^6$ SPH particles, the disk extended from $R_{\rm in} = 5$ au to $R_{\rm out} = 300$ au, with reference radius $R_0$ = 100 au, and a surface density profile was evaluated according to

\begin{equation}
    \Sigma(R) = \Sigma_0 \left( \frac{ R}{R_0} \right)^{-p} \exp \left[\left(\frac{R}{R_0} \right)^{(2-q)} \right]\left( \-\sqrt{\frac{r}{R_{\rm in}}} \right).
\end{equation}

The density normalization $\Sigma_0$ = 1.93 g cm$^{-2}$ was chosen to set an initial total gas mass of 0.04 M$_\odot$ (compatible with the initial disk mass reported by \citealt{Pinilla15} and \citealt{Miley19}). The power-law index, $p$, is equal to 1, while the index q refers only to the sound speed. In the vertical direction, the volume density has a Gaussian profile with a Gaussian width equal to $H/R=(H_0/R_0)(R/R_0)^{1/2-q}$, where we fixed $H/R=0.07$ at $R = R_0$, and $q=0.25$.
The gas-to-dust mass ratio was initially set to 100 (hence, the initial dust mass was 4$\times 10^{-4}$ M$_\odot$), and gas and dust initially had the same surface density profile.
After a few orbits of the planets, the dust settles and forms a layer in the midplane with thickness H$_{\rm d}$ given by \citep{Fromang2009}: $H_{\rm d} = H_{\rm g} (\alpha_{\rm SS}/(St + \alpha_{\rm SS}))^{1/2}$, where $H_{\rm g}$  is the gas disk height and $\alpha_{\rm SS}$ is the \citet{ShaSun77} viscosity. Two different dust-sized particles were included to study the behavior of the small and large dust components, $a_{1}$ = 1.0\,\micron\, and  $a_{2}$=1.0\,mm. The intrinsic grain density was $\rho_{d}= 1$ g cm$^{-2}$, and the initial dust mass was 0.67$\times 10^{-4}$ M$_\odot$ and 3.3$\times 10^{-4}$ M$_\odot$ for the small and large dust particles, respectively. With these assumptions, the initial Stokes number ($St = \rho_d a_{1,2}/\Sigma_{\rm gas}$) was smaller than 1 throughout the whole disk. 

The central star had a mass of 2.4 M$_\odot$; a first giant planet with mass M$_{1,0} = 3$ M$_{\rm J}$ was located inside the inner cavity at a distance of R$_{1,0} = 15$\,au, and a second planetary mass companion with initial mass M$_{2,0} = 1$ M$_{\rm J}$ was initialized in the outer part of the disk, R$_{2,0} = 
170$\,au, on slightly eccentric orbits with eccentricity $e=0.2$. The star and the two planets were represented using sink particles \citep{Bate95}. The mass accretion radius was set to one-fourth of the Hill radius for the planets and 5\,au for the star (as we were not interested in modeling the innermost region of the disk, 
this assumption does not affect the results of the simulation).
The SPH artificial viscosity parameter, $\alpha_{\rm AV}$, was set to=0.2 in order to have an effective \citet{ShaSun77} viscosity of $\alpha_{\rm SS} $ = 0.005.  
The time unit corresponds to the orbital period of the outer planet, T$_{\rm orb}=1880$ yr.
The simulation was stopped after 50 T$_{\rm orb}\sim 10^5$ yr. This timescale is large enough to investigate the spiral structures driven by the outer planet \citep{Dong2016,Rosotti19} and the interactions between the planets and the disk.

\begin{figure*}
\centering
   \includegraphics[width=0.9\textwidth]{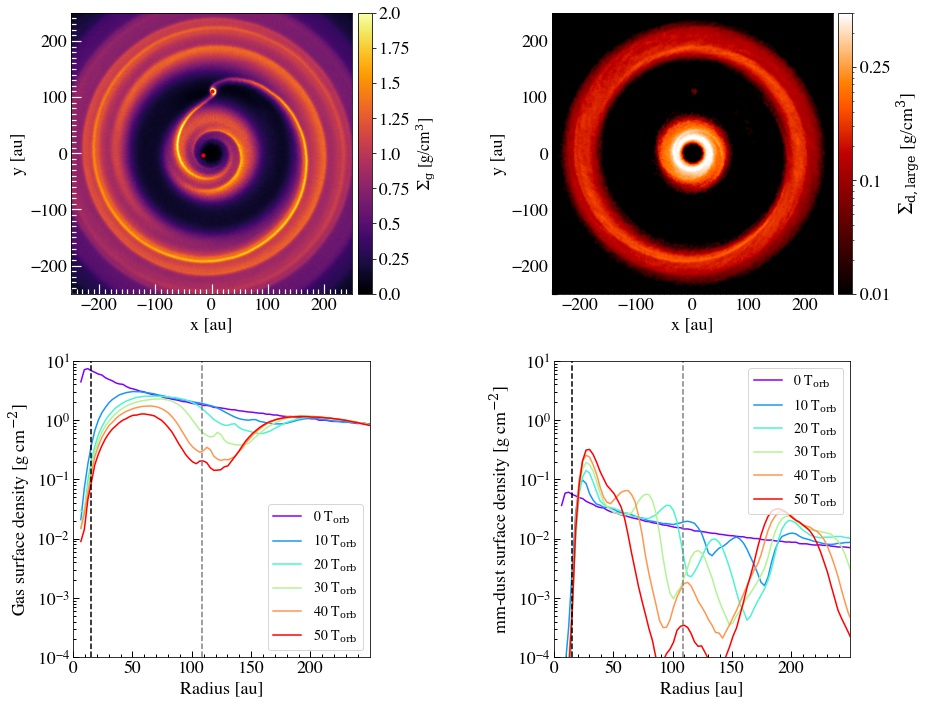}
      \caption{SPH results. ({\it top}) Surface density of gas and 1\,mm dust grains after $t=50\,{\rm T}_{\rm orb}$. The positions of the planets are shown as red dots in the left panel. ({\it bottom}) Temporal evolution of the azimuthally averaged surface density profiles. The final positions of the two planets are indicated by the vertical dashed lines. The shape of the outer gap, both in gas and in dust, reflect the radial migration of the planet.}\label{fig:4simulations}
      \end{figure*}

\subsection{SPH results: Surface density evolution}
\begin{figure*}
    \centering
    \includegraphics[width=0.9\textwidth]{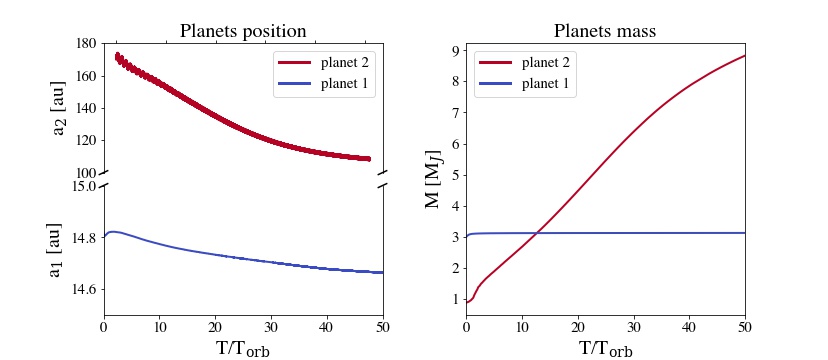}
    \caption{SPH results: temporal evolution of the orbital radius (left) and mass (right) of the two planets.}
    \label{fig:phantom2}
    \end{figure*}

The results of the simulations are shown in Fig.~\ref{fig:4simulations}: The top panels show the surface density
distribution of the gas ($\Sigma_{\rm g}$) and of the large dust grains ($\Sigma_{\rm d, large}$) at $t = 50\,{\rm T}_{\rm orb}$. The bottom panels show the time evolution of the azimuthally averaged surface density profiles. 
The inner planet remains close to its initial position, $a_{i}\sim$ 15 au, opening a gap in the dust and in the gas with a sharp rim at $\sim 20$\,au. The outer planet carves a gap in the gas and in the dust while migrating from its initial orbit to a final orbital radius of $a_{o} \sim 110$\,au. The outer planet decreases its migration speed after about 40 T$_{\rm orb}$; at the end of the simulation (t = 50 T$_{\rm orb}$), when migration has almost stopped, the gas gap extends from $\sim 80-130$\,au while the large dust gap is wider (from $\sim 40$\,au to $\sim$ 150\,au) and deeper than the gas one. The interaction between the outer planet and the disk gives rise to a spiral pattern inward and outward of the planetary orbit.
The large dust grains are confined in two pressure maxima that are induced by the presence of the two planets: The inner one creates the thick bright inner ring ($20-40$\,au), accumulating all the dust that is drifting from the $R < 110$\,au region, while the outer planet traps all the dust that is drifting from the outermost part of the disk in the second pressure maxima located at $\sim 200$\,au. The presence of the spiral arms leaves a signature in the dust distribution: Regions of dust over-density are visible where the spiral arm crosses the dust rings. 
At the end of the simulation, the two planets have a mass of M$_{1} = 3.1 M_{\rm J}$ and M$_{2} = 8.5 M_{\rm J}$, respectively. We note that the mass of the outer planet is just 1.4 times larger than that estimated in Sect.~\ref{sec:results}. Figure~\ref{fig:phantom2} shows the temporal evolution of the orbital radius and mass of the two planets in the simulation: The mass accretion rate on the inner planet drops after about 10 T$_{\rm orb}$, and the accretion rate on the outer planet starts to flatten after about 40 T$_{\rm orb}$; the outer planet mass is about 8.5 $ M_{\rm J}$ at t= 50 T$_{\rm orb}$. Extrapolating the curve of the mass accretion rate, the final mass of the outer planet is expected to be nearly 10 $M_{\rm J}$. During migration, the gas-planet interaction damps the eccentricity of the outer planet, reducing it from 0.2 to 0.05 after 50 orbits. As a consequence, the outer dusty ring is slightly eccentric (e $\sim$ 0.05). Interestingly, in our PHANTOM simulations, when the initial eccentricity of the planet is set to zero (not shown here), the eccentricity of the outer planet grows to 0.05 and 0.1 at $t=50~T_{\rm orb}$. The final gas and total dust masses are 2.8$\times 10^{-2}$ M$_{\odot}$ and 4$\times 10^{-4}$ M$_{\odot}$, respectively. 

\subsection{SPH results: Synthetic images}
\begin{figure*}
   \centering
    \includegraphics[width=8cm]{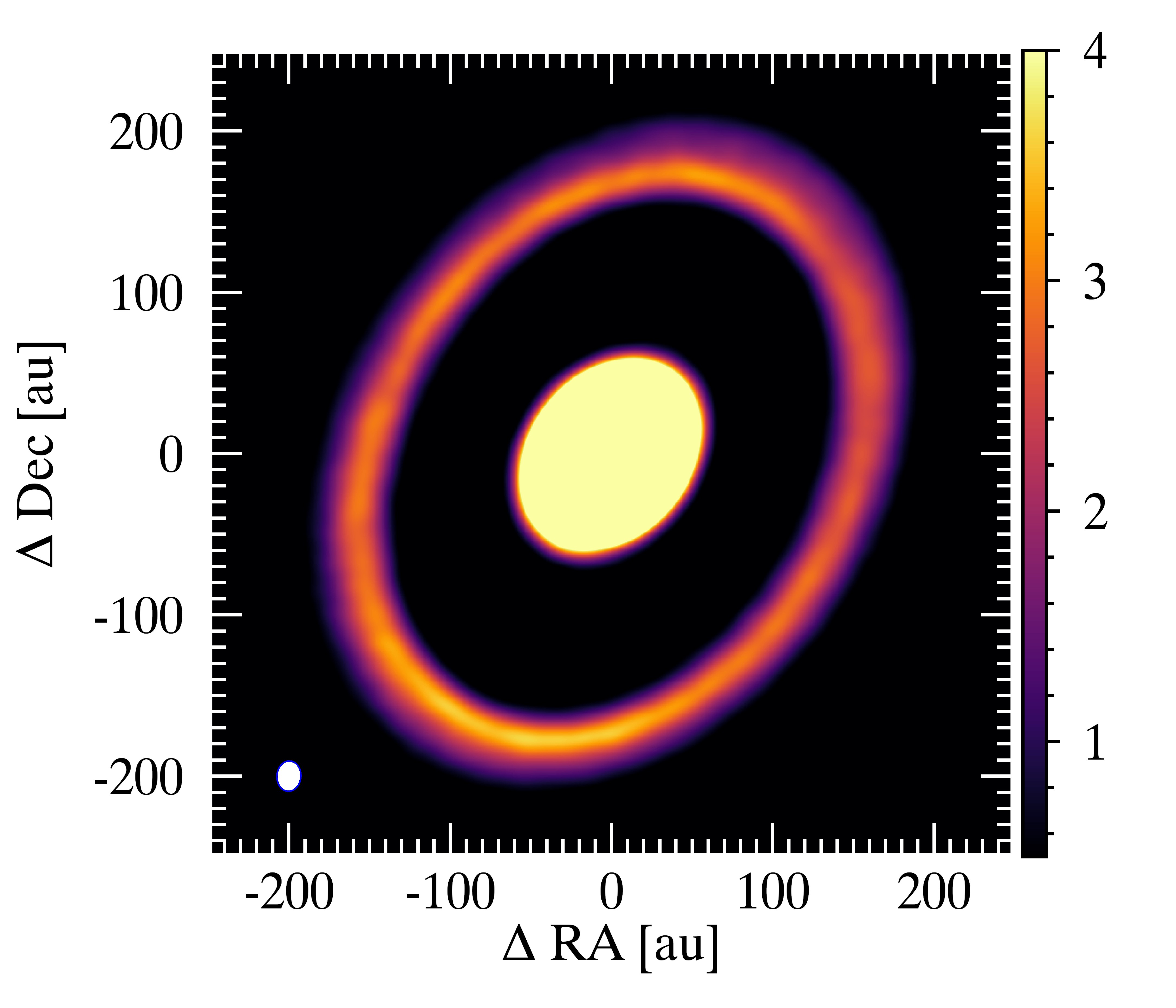}
    \includegraphics[width=8cm]{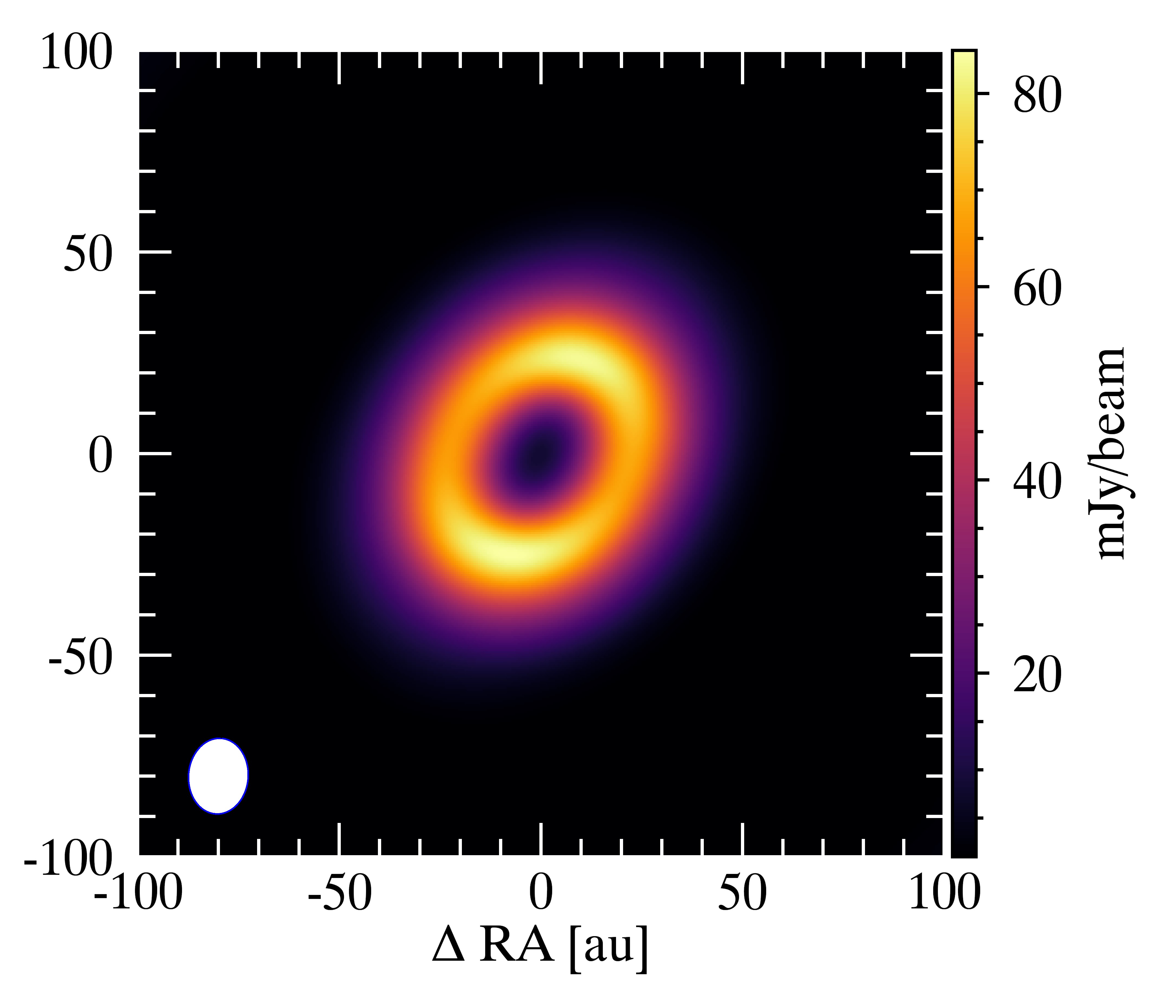}
      \caption{Synthetic ALMA 870\,\micron\    continuum images based on the SPH simulations. ({\it left}) Zoom in on the emission from the inner ring convolved with a beam of $0\farcs036 \times  0\farcs027$ matching the angular resolution of data set A. ({\it right}) Synthetic image convolved with a beam of $0\farcs156 \times 0\farcs127$, similar to data set A+B.}
         \label{fig:mcfost}
   \end{figure*}

Synthetic observations are computed with MCFOST \citep{Pinte06,Pinte09}. A Voronoi tessellation maps each SPH particle to an MCFOST cell without interpolation. The DIscANAlysis (DIANA) dust opacity \citep{Woitke16} was adopted, assuming a fixed dust mixture composed of 70$\%$ silicate and 30$\%$ amorphous carbonaceous and a grain size distribution with slope of -3.5 in a range from 0.03 $\mu$m to 3 mm. The expected emission maps were computed using ray tracing and 10$^{8}$ photon packets, with a disk inclination of $i$ = 42$^{\circ}$, a distance $d = 110$\,pc, and PA = 147$^{\circ}$. The stellar spectral model is selected from the stellar isochrones from \citet{Siess2010}. Finally, the synthetic 870 $\mu$m continuum images were produced with the \textsc{casa} tasks \textsc{symobserve} and \textsc{symanalyze}, adopting the array configuration 4.9 that almost matches the achieved beam of the concatenated data set.
Figure~\ref{fig:mcfost} shows the synthetic dust continuum images at 870\,\micron ~for the $50\,{\rm T}_{\rm orb}$ snapshot.

The synthetic images reproduce several observational findings, such as the double ring-like structure with the inner ring ($\sim$ 20-40\,au) much brighter than the outer one ($\sim$ 150-250\,au), the wide dust gap ($\sim$ 40-150\,au), and the 
asymmetry in the outer disk. 
As noted in Sect.~\ref{sec:results}, there is a tentative detection of a point-like source at nearly 100\,au from the star (S4). Further observations are needed to confirm the detection and unveil the nature of such emission.

\subsection{Simulation caveats}
While the synthetic dust continuum images reproduce several of the observational results, the actual flux level of the outer ring is not recovered as it is overestimated by an order of magnitude. This may be due to the assumptions on the physical disk setup and, in particular, to the slope of the surface density profile or to the uncertainties on the radiative transfer parameters, such as the opacity of the dust, its chemical composition, and the boundary values of the grain size distribution; the emission at 870 ${\mu}$m becomes fainter for higher values of $a_{\rm max}$ or for dust composed of silicates only. On the other hand, a larger initial disk size ($R_{\rm out} > 300$\,au) and a steeper surface density profile
($p$ > 1) could produce a fainter outer disk.
   
\section{Discussion}\label{sec:discussion}
\subsection{Disk-planet interaction}
The analysis presented above hints at the presence of two giant protoplanets at $\sim 15\,$au and $\sim 110\,$au. The planets
shape the disk structure and dynamics, trapping the large dust grains in two distinct rings: The outer planet traps the dust 
in a ring-like structure centered at 200\,au while the inner planet prevents the inward dust drift, blocking the large grains at 
$r \sim 20-40\,$au. Dust trapping in multiple rings was proposed by \citet{Pinilla15a} for the general case of multiple giant
protoplanets embedded in a protoplanetary disk and by \citet{Pinilla15b} in the specific case of HD 100546. A similar configuration  has been found in other disks, such as HD 169142, where two giant protoplanets at $\sim 18\,$au and $\sim 51\,$au give rise to a 
double-ring dust-like structure \citep[e.g.,][]{Toci20}.
In the case of HD 100546, the presence of the outer massive planet perturbs the gas structure and dynamics, inducing a two-armed 
spiral in the $R < 100$~au part of the disk, similar to the spiral pattern observed in the inner disk region in the near-infrared 
by, for example, \citet{Garufi16} and \citet{Follette17}. No substantial misalignment has been found between the inner and outer disk (Sect.~\ref{sec:results}).
The gas spiral arms in the inner disk can also explain the gas dynamics inward of 100\,au observed by \citet{Pineda14}, \citet{Walsh15}, and \citet{Pineda19}. 
Further ALMA observations of CO isotopologs at medium angular resolution ($\sim 0\farcs15$) could potentially reveal the gas spiral structure.

The SPH simulations presented here suggest that the outer planet may have been formed farther out in the disk (the initial position is 170\,au) 
and may have subsequently migrated inward. Other protoplanetary disks show the imprints of giant protoplanets on wide orbits ($R \gtrsim 100$~au), 
such as HD 163296 \citep[e.g.,][]{Isella16,Isella18,Pinte18}, HD 97048 \citep{vanderPlas16, Pinte19}, and AS 209 \citep[e.g.,][]{Fedele18, Guzman18, Favre19}.
The formation of (massive) planets at such a large distance from the star is at odds with the standard core-accretion model and instead favors 
the gravitational instability scenario \citep{Kratter16}. 

\subsection{Dust gap and water emission}
The presence of the $40 - 150\,$au dust gap has some important astrochemical implications regarding the detection of water ice and water vapor from the disk around HD 100546.
\citet{Honda16} reported an absorption feature at 3.1\,\micron\ that is likely due to water ice grains. 
The ice absorption band arises from a region between nearly 40 and 120\,au from the star.  Far-infrared spectroscopy with the Heterodyne Instrument for the Far-Infrared (HIFI) onboard of Herschel has also revealed emission of the ground-state H$_2$O lines \citep{Du17,vanDishoeck21}. The velocity profile of the two water lines indicates that the inner radius of the water emission is $\sim40\,$au, while the outer radius is poorly constrained as the HIFI spectra are spatially unresolved. Both features (ice absorption and gas emission) are in agreement with photo-desorption models \citep{Honda16,vanDishoeck21}.
The spatial coincidence between the millimeter dust gap, the water ice absorption, and the ground-state water vapor emission is striking: Because of the lower dust extinction in the dust gap, the stellar UV photons can reach the colder layers and evaporate water molecules via UV photo-desorption. This implies that the dust gap is not fully void of dust, that there must be a residual of small dust grains. 

\section{Conclusions}\label{sec:conclusions}
This paper has presented a new analysis of archival ALMA 870\,\micron ~images, revealing, in addition to the previously known inner cavity within 20 au, a wide dust gap between $r \sim 40-150\,$au and a previously unresolved faint dust ring beyond 150\,au.
The emission in the outer disk is not homogeneous, and two main substructures are detected, peaking at $\sim 180\,$au and $\sim 220\,$au, respectively. The azimuthally asymmetric emission is likely caused by an eccentric ring with eccentricity $e = 0.07\pm 0.01$.
The observational findings and the comparison with hydrodynamical 
simulations hint at the presence of two giant protoplanets at nearly 15\,au and 110\,au, with masses of $\sim 3.1\,M_{\rm J}$ and 8.5$\,M_{\rm J}$, 
respectively. 
The massive outer planet challenges current formation models via core accretion.


\begin{acknowledgements}
This work was supported by the PRIN-INAF 2019 Planetary Systems At Early Ages (PLATEA).
CT and GL have received funding from the European Union's Horizon 2020 research and innovation programme under the Marie Skłodowska-Curie grant agreement No 823823 (DUSTBUSTERS RISE project).

\end{acknowledgements}

\begin{appendix} 
\section{\textsc{Galario} fit results}
The geometrical properties of the 870\,\micron ~ continuum emission were obtained by fitting the visibility of data set B with
\textsc{galario} \citep{Tazzari18} and \textsc{emcee} \citep{Foreman13} with uniform priors. The brightness profile is the sum of two Gaussian rings with the same inclination, position angle, and phase center:

\begin{equation}
    I(R) = I_1 e^{-(R - R_1)^2/2\sigma_1^2} + I_2 e^{-(R - R_2)^2/2\sigma_2^2}
.\end{equation}

\begin{figure*}
    \centering
    \includegraphics[width=18cm]{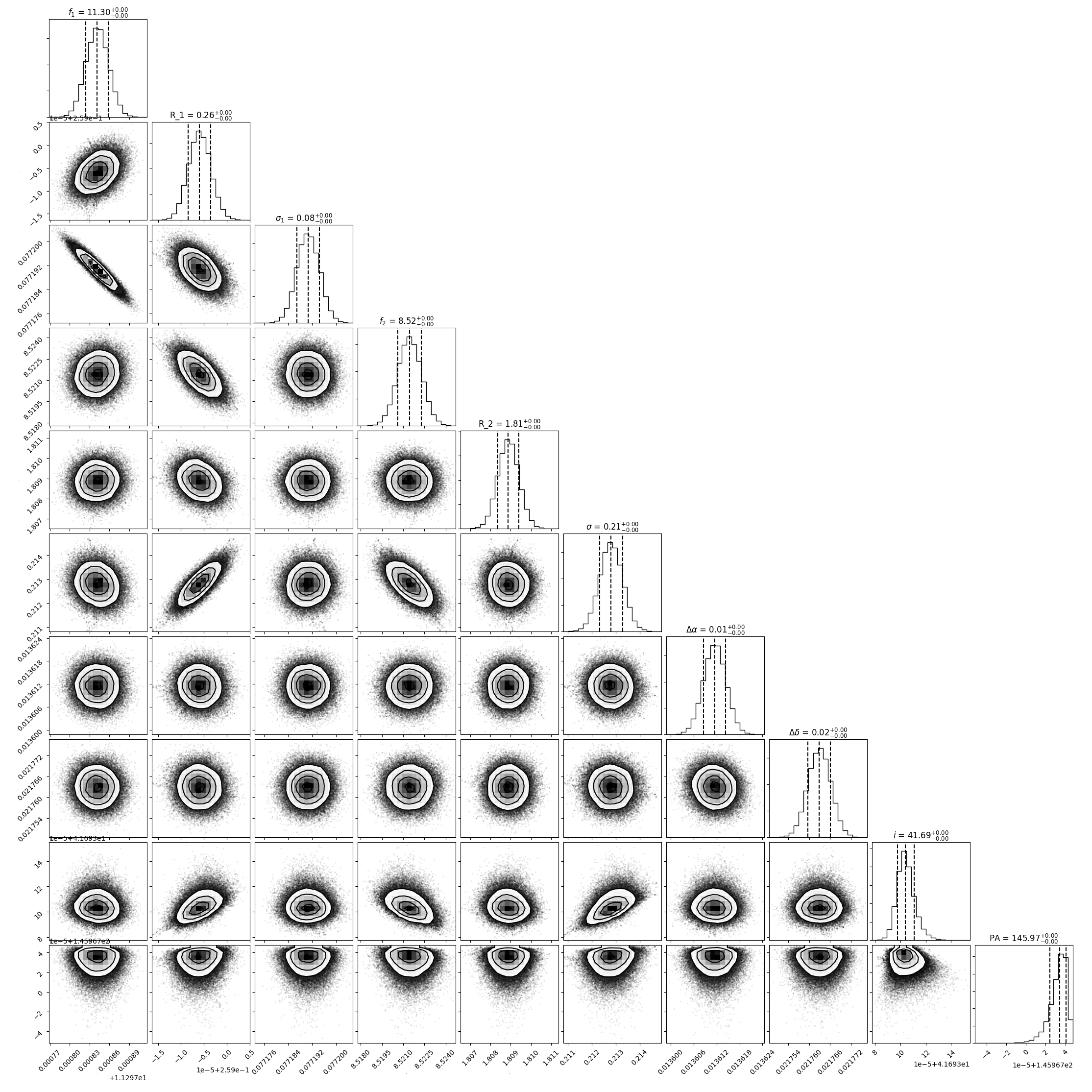}
    \caption{\textsc{Galario} fit results: 1D and 2D marginalized distributions of the posterior sampling obtained with MCMC.}
    \label{fig:corner}
\end{figure*}

The best-fit model parameters are found by minimizing the $\chi^2$:  

\begin{equation}
\chi^{2} = \sum_{j=0}^{N}|V_{\mathrm{obs}}(u_{j}, v_{j})-V_{\mathrm{mod}}(u_{j}, v_{j})|^{2} w_{j}\,,
\end{equation}where $w_{j}$ is the weight of the observed $(u_{j},v_{j})$ visibility points. The posterior distribution is computed as $\exp(-\chi^2/2)$. \textsc{Emcee} is initialized with 50 walkers and with 40000 Markov chain Monte Carlo (MCMC) steps". The modeling results are listed Table~\ref{tab:galario} along with the initial range of the prior functions. The best-fit values correspond to the 50\% percentile of the marginalized distributions, and the uncertainties correspond to half the interval between the 16\% and 84\% percentiles. The percentiles were computed by reading every ten steps. Figure~\ref{fig:corner} shows the 1D and 2D marginalized posterior distributions.

\begin{table}[]
    \centering
    \caption{\textsc{Galario} modeling results. Columns ``min'' and ``max'' represent the range of the uniform priors. }
    \begin{tabular}{llccc}
    \hline\hline
        Parameter & Unit & Min  & Max  &  Best-fit\\
        \hline
        log I$_1$   & [Jy/beam]     & 9     & 12   &  11.297 $\pm$ 0.001   \\
        R$_1$       & [\arcsec]     & 0.2   & 0.35 &   0.258 $\pm$ 0.001   \\
        $\sigma_1$  & [\arcsec]     & 0.05  & 0.10 &   0.077 $\pm$ 0.001   \\
        log I$_2$   & [Jy/beam]     & 7     & 9    &   8.521 $\pm$ 0.003   \\
        R$_2$       & [\arcsec]     & 1.7   & 2.1  &   1.808 $\pm$ 0.004   \\
        $\sigma_2$  & [\arcsec]     & 0.2   & 0.4  &   0.212 $\pm$ 0.001   \\
        $i$         & [$^{\circ}$]  & 40    & 60   &  41.693 $\pm$ 0.047   \\
        PA          & [$^{\circ}$]  & 130   & 160  & 145.967 $\pm$ 0.010   \\
        $\Delta\alpha$ & [\arcsec]  & -0.05 & 0.05 &  0.0136 $\pm$ 0.001   \\
        $\Delta\delta$ & [\arcsec]  & -0.05 & 0.05 &  0.0217 $\pm$ 0.001   \\
    \hline\hline
    \end{tabular}
    \label{tab:galario}
\end{table}

\end{appendix}

     \bibliographystyle{aa} 
     \bibliography{mybib} 

\end{document}